Aluminum doping improves the energetics of lithium, sodium, and magnesium storage in silicon

*Fleur Legrain[1] and Sergei Manzhos[1]\**

[1]Department of Mechanical Engineering, National University of Singapore, Block EA #07-08, 9 Engineering Drive 1, Singapore 117576, Singapore

*\*E-mail:* mpemanzh@nus.edu.sg*;  Fax: +65 6779 1459; Tel: +65 6516 4605*

ABSTRACT. While Si is an effective insertion type anode for Li-ion batteries, crystalline Si has been shown to be unsuitable for Na and Mg storage due, in particular, to insufficient binding strength. It has recently been reported that Si nanowires could be synthesized with high-concentration (several atomic %) and dispersed Al doping. Here we show based on density functional theory calculations that Al doping significantly improves the energetics for Na and Mg insertion, specifically, making it thermodynamically favored versus vacuum reference states. For high Al concentrations, the energy of Mg in Al-doped Si approaches the cohesive energy of Mg. However, the migration barriers for the diffusion of Li (0.57-0.70 eV), Na (1.07-1.19 eV) and Mg (0.97-1.18 eV) in Al-doped Si are found to remain about as high as in pure Si, likely preventing effective electrochemical sodiation and magnesiation.

1. **Introduction**

The progress towards renewable but intermittent sources of electricity as well as the development of electrical vehicles calls for efficient energy storage systems [1]. Electrochemical batteries which offer relatively high energy densities have been attracting renewed interest [2]. Among them, the Li-ion batteries are already of widespread use, especially for portable electronics. However, because of safety issues and the limited lithium resources, alternatives are needed [3]. Na and Mg, which are abundant, cheap, and benign, have recently attracted much scientific attention as post-Li storage systems [4-6]. Among the remaining issues for their commercialization, a suitable anode material, enabling the insertion/de-insertion of the metal atoms at a reasonable rate, is still highly demanded. Si, which provides relatively high theoretical specific capacities for Na (~950 mAh $g^{-1}$) and Mg (~3820 mAh $g^{-1}$) storage and which was already shown to be practical for Li (theoretical capacity of ~4200 mAh $g^{-1}$ and demonstrated capacities of >1000 mAh$g^{-1}$ for thousands of cycles [7]), was an obvious candidate. But, as other good anode materials for Li (e.g. graphite [8]), diamond Si was found to not work for Na and Mg [9-11]. First-principles calculations have shown that while the final states of charge (NaSi and $Mg_2Si$) have negative heats of formation [12], the insertion of Na and Mg in pure Si is thermodynamically unfavored (for small Na and Mg concentrations) [13-16].

DFT (density functional theory) calculations have also shown that modification of the ideal pure crystalline Si, especially amorphization [14, 17], could make Si suitable for the insertion of Na and Mg. Specifically, Al is a promising dopant to improve the performance of battery electrodes. For example, Al-C nanoclustered anodes [18] and Al-doped $Li_4Ti_5O_{12}$ anodes [19] have been shown to exhibit significantly better cyclability and capacity compared to pure host materials. Similarly, experiments on Al-doped Si [20, 21] and eutectic Al-Si (~88:12 wt%) [22] have also been reported, and have shown enhanced cyclability and capacity over pure Si and Al anodes, respectively.

Moutanabbir et al. have reported that Al concentrations which exceed by orders of magnitude the equilibrium solid solubility (i.e. average Al concentrations across samples as high as 4.3 at. %) are achievable by growing Si nanowires on a catalyst [23]. In this work, we show that doping Si with Al at concentrations of this magnitude improves significantly the energetics for Na and Mg (as well as Li) storage in Si.

2. **Methods**

A 64 atom cell was used to model Si and Al-doped Si. The electronic structure was computed using DFT [24] and the SIESTA code [25]. The PBE exchange-correlation functional [26] and the DZP basis set (double-ξ polarized orbitals) were used. A cutoff of 100 Ry was used for the Fourier expansion of the density. Core electrons were modeled with Trouiller-Martins pseudopotentials [27]. The basis sets of Si, Al, Li, Na and Mg were tuned to reproduce their cohesive energies (see supporting information). The calculated and (vs.) the experimental values of $E_{coh}$ (adjusted for the effect of zero-point motion, ZPE, where for Li, Na and Mg, the ZPE corrections are computed to be 0.034 eV, 0.016 eV and 0.029 eV, respectively) give for Si: 4.66 vs. 4.68 eV by using a ZPE correction of 0.06 eV [28, 29], Al: 3.51 vs. 3.43 eV [30], Li: 1.67 vs. 1.66 [31], Na: 1.14 vs. 1.13 eV [31], Mg: 1.55 vs. 1.54 eV [31]). Geometries were optimized until forces on all atoms were below 0.01 eV/Å and stresses below 0.1 GPa. Brillouin-zone integrations were done with a 3×3×3 k-point Monkhorst-Pack mesh [32]. Unrestricted spin-polarized calculations were performed, but spin polarization was found insignificant in bulk.

The insertion energetics of Al in Si and of metal atoms in pure and Al-doped Si were analyzed by computing the defect formation energies ($E_f$)

$$E_f = (E_{mSi/nAl} - mE_{Si} - n_{Al}E_{Al})/n_{Al}$$

$$E_f = (E_{X/nM} - E_X - n_M E_M)/n_M$$

where $E_{mSi/nAl}$ is the total energy of the simulation cell consisting of $m$ Si atoms and $n_{Al}$ Al atoms; $E_{Si}$ is the energy of a Si atom in pure Si (i.e. Si bulk); $E_{Al}$ is the energy of an Al atom in vacuum (Al bulk) for a vacuum (bulk) reference state; $E_{X/nM}$ is the total energy of the simulation cell with $n_M$ metal atoms M (M = Li/Na/Mg) inserted; $E_X$ is the energy of the cell X without alkali metal atoms (X designating $Si_{64}$, $Al_1Si_{63}$, $Al_2Si_{62}$, $Al_4Si_{60}$ or $Al_8Si_{56}$); $E_M$ is the energy of a metal atom M in vacuum or bulk, giving the defect formation energies versus vacuum or bulk reference states, respectively.

Diffusion barriers were computed for one Li, Na, and Mg atom in the supercell by constrained optimization, in which the dopant atom's projection on the line connecting the initial and the final sites of the diffusion step was fixed and stepped. A step of 0.3 Bohr (~0.16 Å) was used, which corresponds to 15 images per diffusion path. The atoms farther than 5 Å from the initial and final sites were fixed in the diffusion direction to avoid translation of all atoms. The cell was fixed. All other degrees of freedom were relaxed.

3. **Results and discussion**

We first study the insertion of Al in diamond Si and the three following insertion defects are considered: the substitutional (S), and the tetrahedral (T) and hexagonal (H) interstitial defects (see Figure 1 (a), (b), and (c), respectively). For one Al dopant in a 64 Si atom cell (corresponding to an Al concentration of ~1.6 at. %), the corresponding defect formation energies are given in Table 1. For all insertion defects, the defect formation energies are found positive versus the bulk reference state, i.e. relative to the cohesive energy of Al. This means that at this concentration, the insertion of Al in Si is unfavored compared to Al clustering, which is expected given the lower value of solid solubility of Al in Si [33]. Among the three defects considered, the substitutional defect is found the most preferred by more than 2 eV. At a concentration of ~1.6 at. % the S defect is therefore predominant and Al atoms are mainly located at Si sites. The insertion of 2, 4 and 8 Al dopants, corresponding to a concentration of ~3.1, ~6.2 and ~12.5 at. %, respectively, is also considered, and the Al atoms are inserted by maximizing the inter-dopant distances (see Figure 1 (d)). These Al concentrations are practically achievable, as they are of the same order of magnitude as the one observed in the Al-doped nanowires synthesized by Moutanabbir et al (~4.3 at. %) [23]. The well-separated Al configurations are chosen because in Ref. [23], the Al impurities were found to be homogeneously distributed in the nanowire and to not form precipitates or clusters. The computed defect formation energies (Table 2) show that the tetrahedral interstitial site starts (slightly) to be favored when 8 dopants are inserted, but that the substitutional defect remains preferred up to a concentration of 4 atoms per simulation cell, which includes the concentration reported by Moutanabbir et al. (~4.3 at. %). We considered alternative distributions of Al in Si as well as possibilities of diffusion of Al in Si, and the changes in defect formation energies were found to be unimportant for different distributions of Al dopants, and Al diffusion in Si was found to be unlikely (see supporting information).

We focus here on the effect of Al doping on the Li, Na, and Mg storage properties of Si. The results are summarized in Table 3. The defect formation energies of Li, Na, and Mg of -1.37 eV, 0.37 eV, and 0.64 eV, respectively, in non-doped Si are in good agreement with previously reported values, ranging -1.37...-1.42 eV for Li, 0.37…0.75 eV for Na, and 0.32…0.64 for Mg [13-15]. Specifically, the defect formation energies for the insertion of Na and Mg are positive versus both vacuum and bulk metal references states [13-15]. Therefore, diamond Si does not thermodynamically favor the insertion of Na and Mg. The computed (here and elsewhere [13, 34]) $E_f$ of Li in pure Si is also somewhat (by ~0.3 eV) weaker than Li's $E_{coh}$, although electrochemical insertion of Li in Si is known to be efficient [35]. It is known, however, to proceed via the movement of a lithiated front rather than single-atom diffusion [7, 36]. The defect formation energies for the insertion of Li, Na, and Mg atoms in Al-doped Si for all Al concentrations mentioned previously (~1.6, ~3.1, ~6.2 and ~12.5 at. %) are then computed. Since the T site was reported to be the equilibrium insertion site in Si, we insert one metal (M) atom at the three nearest ($T_1$, $T_2$, $T_3$) and at the farthest ($T_f$) T sites from an Al dopant (see Figure 1 (e)). From the results for all configurations, it appears that the most preferred sites minimize the Al-M bonds and maximize the number of nearest Al neighbors. For (the concentrations corresponding to) 1 and 2 Al atom(s) inserted, the preferred site is found to be the nearest site $T_1$ (~2.4 Å) and the defect formation energies are found to decrease (i.e. insertion of Li, Na and Mg stabilized) as the Al-M bond distance is decreased, the dependence being more significant for Mg > Na > Li. For 4 Al atoms and for Li and Mg only, the most stable site is found to be the third nearest site $T_3$ (~4.8 Å), likely because in that configuration, the metal atom has 4 equivalent nearest Al neighbors which stabilize it. For 4 Al atoms and for Na, the most preferred site remains the nearest site $T_1$ (~2.4 Å). We also investigated, by computing the strain energies, whether this difference in behavior is due to a larger stress generated by Na in $T_3$ [37] (see supporting information), but results have shown that the effect of stress on the energies is not important. For 8 Al atoms and for all metal atoms, the most stable site is the second nearest site $T_2$ (~2.8 Å), in which the metal atom has two equidistant nearest Al neighbors. All Li, Na, and Mg appear to be stabilized when surrounded by Al dopants.

The lowest defect formation energies at all Al concentrations studied are given in Table 3. The results show that Al doping improves the thermodynamics of insertion of all Li, Na, and Mg atoms in Si. While non-doped Si gives positive defect formation energies (i.e. unfavored insertion) versus bulk and vacuum reference states for Na and Mg (and versus bulk only for Li), Al doping at a concentration of only ~1.6 at. % (i.e. 1 Al) provides negative defect formation energies (i.e. favored insertion) versus vacuum for Na and Mg (and versus vacuum and bulk for Li).

Al doping at higher concentrations stabilizes even more the insertion of Na and Mg: the defect formation energies decrease as the Al concentration is increased. The stabilization provided by Al doping is found more significant for Mg > Na > Li. Al concentrations of ~1.6/~3.1/~6.2/~12.5 at. % (1/2/4/8 Al inserted) stabilize the insertion of Li, Na and Mg by 0.75/0.91/1.00/1.06, 0.79/0.98/1.00/1.12 and 1.10/1.89/2.16/2.42 eV, respectively. It has been experimentally shown that Si could be doped with Al at concentrations of that order of magnitude (~4.3 at. % in Ref. [23]). At the specific concentration of ~4.3 at. %, the defect formation energies obtained by interpolation are -2.31 (-0.64), -0.62 (0.52) and -1.35 (0.20) eV versus vacuum (bulk) reference states for the insertion of Li, Na, and Mg, respectively. Therefore, Al doping improves significantly the insertion energetics of all Li, Na, and Mg and could potentially allow for the insertion of Na and Mg in Si.

We use here the densities of states and the Mulliken [38] charges to investigate the stabilization mechanism of Li/Na/Mg insertion in Si with Al doping. The densities of states (see supporting information) show that when Li/Na/Mg inserts into pure Si, the Fermi level moves from the band gap to the conduction band. It has been established that the valence electron(s) of the metal atom go(es) into an anti-bonding state in the Si conduction band [13, 39]. However, doping Si with Al makes the Fermi level move from the band gap to the valence band where Al states appear, which attract electrons. This is also confirmed by the negative Mulliken charge on Al (and positive on Si) of -0.74 |$e$|. When Li/Na/Mg inserts into Al-doped Si, the electrons donated by the alkali atoms [13, 39] then fill first the empty bonding states in the valence band created by Al doping before going into the conduction band (if there are more electrons than holes). Each Al dopant seems to form one hole, since the insertion of one Mg (which brings two valence electrons) in Si doped with one Al puts the Fermi level in the conduction band, while the insertion of one Li, Na (Mg) in Si doped with 1 (2 for Mg) Al moves the Fermi level to the band gap, and the insertion of one Li, Na (Mg) in Si doped with higher Al concentrations moves the Fermi level to the valence band. The Al-Si bonds are also found to be longer (2.45 Å) than the Si-Si bonds (2.39 Å). When an alkali metal atom inserts at the $T_1$ site in Al-doped Si (1 Al), Li/Na/Mg atom gives its *s* electrons, more exactly, 0.55/0.35/1.35 *e*, to the Si framework (we do not count the occupation of basis functions with *p* character towards the alkali atom charge, as their occupation *is* due to charge redistribution towards Si). The charge donations are comparable to those found in pure Si: 0.54/0.36/1.30 *e*. The electronic charge received by the Al dopants is decreased upon insertion of a metal atom (from 0.74 *e* to 0.56/0.59/0.55 *e* for Li/Na/Mg), and to a larger extent the larger the charge given by the alkali metal atom. It appears that the electronic charge given by the metal atom helps strengthen the Al-Si bond in the direction of the metal atom, as shown by the bond lengths: the preferred Al-Si bond has a length of 2.40/2.39/2.39 Å for bonds near Li/Na/Mg (vs. a distance of 2.49/2.53/2.54 Å for the three other Al-Si bonds).

The insertion energetics of 2, 4, and 8 Li/Na/Mg is also investigated in Al-doped Si, for the same Al concentrations (~1.6, ~3.1, ~6.2, ~12.5 at. %) and T sites (the three nearest -$T_1$, $T_2$, $T_3$- and the farthest – $T_f$- T sites from an Al dopant). Among all possible configurations, the metal atoms are inserted by maximizing their inter-atomic distances, since (i) Li, Na and Mg have been reported to repel each other in Si [13, 15] (ii) all the metal atoms are equivalent (i.e. same surrounding) for the same concentration of M and Al atoms. The defect formation energies for all the sites and the Al and metal concentrations considered appear in Figure 2. As previously reported [13, 15, 34] for pure Si, the defect formation energies increase (i.e. insertion destabilized) as the alkali metal concentration is increased from $x = 1/64$ to 1/8, with *x* the number of metal atoms per Si (or Al) atom. This destabilization due to the increase in metal concentration happens also for all Al concentrations considered. But Al doping helps to stabilize the insertion at all the M concentrations, and for an Al concentration of ~6.2/~12.5 at. %, the highest defect formation energy among all metal concentrations (i.e. $x = 1/8$) is -1.62/-2.34, 0.00/-0.75, 0.01/-0.81 eV for Li, Na and Mg, respectively. High-concentration Al doping should therefore allow for the insertion of Na and Mg in Si.

We also investigate the effects of Al doping on the energy migration barriers of Li, Na and Mg in Si. The migration of Li, Na and Mg in Si happens between two T sites via the H site [13, 15, 34, 37]. The migration barriers computed in pure Si for Li (0.56 eV), Na (1.09 eV) and Mg (0.97 eV) are in good agreement with the literature [13, 15, 34, 37]. In Al-doped Si, concentrations of ~1.6 and ~6.2 at. % are considered, for which respectively the migration pathways $T_1$-$T_2$-$T_4$-$T_5$ and $T_1$-$T_2$-$T_4$ are computed. $T_1$ and $T_2$ designate the nearest two T sites, and $T_4$ and $T_5$ two farther T sites. $T_4$ ($T_5$) is at mid-distance between two Al dopants in a concentration of ~1.6 at. % (~6.2 at. %). The migration pathways are shown in Figure 1 (e), the migration energies are plotted in Figure 3, and the energy barriers are given in Table 4. In Al-doped Si, results show that even though the metal atom is more stable in $T_1$ than $T_2$, the energy barriers from $T_1$ to $T_2$ are the same within 0.02 eV as the ones in pure Si (for both Al concentrations). The same applies from $T_4$ to $T_5$ in ~1.6 at. % (energy differences within 0.03 eV). However, energy barriers are significantly higher though $T_2$ to $T_4$ in ~1.6 at. % (increase of 0.14, 0.08 and 0.21 eV for Li, Na, and Mg, respectively) and to a lesser degree in ~6.2 at. % (increase of 0.10, 0.02 and 0.17 eV for Li, Na, and Mg, respectively). That is, diffusion barriers for Na and Mg diffusion remain high in Al-doped Si.

**Conclusions**

In conclusion, Si doped with Al was investigated as a potential anode for Na and Mg ion batteries. DFT calculations show that while the insertion of Na and Mg is not thermodynamically favored in Si, doping Si with Al improves significantly the energetics for Na and Mg storage. At Al concentrations that were recently shown to be achievable experimentally (of several at. %), the insertion of Na and Mg becomes favored versus the vacuum reference state, and in the case of Mg, becomes competitive with the metal's cohesive energy. Importantly, the stabilization of $E_f$ is achieved everywhere in the material at practically achievable dopant concentrations, not just in the vicinity of the dopant (a trap). Therefore, doping Si with Al could make Si work as an insertion material for Na and Mg storage. The insertion energetics of Li in Si is also improved by Al doping, specifically, making Li defect formation energy become lower (stronger) than Li's $E_{coh}$, which is not the case in pure Si (according to all DFT studies to date [13, 14, 34, 40]). Lithiation by single atom diffusion mechanism (rather than by the movement of a phase boundary [7, 36]) should therefore become possible in Al-doped Si. However, the migration barriers for Na (1.07-1.19 eV) and Mg (0.97-1.18 eV) in Al-doped Si remain high (as they are in pure Si) and do not bode well for insertion kinetics.

**Supporting information**

Effect of different distributions of Al dopants, Al diffusion in Si, strain energies, full states of charge, densities of states, basis set information and headers of the pseudopotentials are given in the supporting information.

**Acknowledgments**

This work was supported by the Tier 1 AcRF Grant by the Ministry of Education of Singapore (R-265-000-494-112).

**Figure Captions**

**Figure 1.** (a), (b), and (c) Atomic configurations for substitutional (S), and tetrahedral (T) and hexagonal (H) interstitial defects, respectively, in the silicon lattice. (d) Location of the Al atoms when 4 (pink *or* purple atoms) and 8 (pink *and* purple atoms) Al are inserted in a 64 atom Si cell. (e) In the presence of an Al dopant, location of the nearest three T sites and farther T sites from the Al atom (labelled 1, 2, 3, 4 and 5). The small blue balls show the migration pathway computed for Li/Na/Mg in Al concentrations of ~1.6 at. % ($T_1$-$T_2$-$T_4$-$T_5$) and ~6.2 at. % ($T_1$-$T_2$-$T_4$), where $T_5$ and $T_4$ are at mid-distance between two Al dopant atoms, respectively. Color scheme: yellow – Si, pink and purple – Al, blue – M (Li, Na or Mg),

**Figure 2.** (a), (b), and (c) Defect formation energies versus the vacuum reference state (in eV) for Li, Na and Mg insertion, respectively, in pure and Al-doped Si. Symbol and color schemes: circle – $T_1$, diamond – $T_2$, triangle – $T_3$, star – $T_f$, square – T, white – 1 M, light grey – 2 M, dark grey – 4 M, black – 8 M (M designating Li, Na or Mg). The dashed lines indicate the cohesive energy of Li/Na/Mg.

**Figure 3.** Migration energy pathways for Li (blue), Na (red) and Mg (green) in pure and Al-doped Si. In pure Si (white filled dots), the migration is computed between two equivalent T sites. In Si doped with Al at a concentration of 1.6 at. % (resp. 6.2 at. %) designated on the plot by color (resp. black) filled dots, the migration is computed between $T_1$ and $T_5$ (resp. $T_4$), $T_1$ being the nearest T site from an Al dopant and $T_5$ (resp. $T_4$) a T site at mid-distance between two Al dopants.

TABLES

Table 1 Defect formation energies (in eV) versus vacuum and bulk reference states for the insertion of one Al dopant in a 64-atom cell of diamond Si.

|        | T     | H    | S     |
|--------|-------|------|-------|
| Vacuum | -0.85 | 0.19 | -2.89 |
| Bulk   | 2.66  | 3.70 | 0.62  |

Table 2 Defect formation energies per dopant atom (in eV) versus vacuum for $n$ = 1, 2, 4 and 8 Al atoms inserted in a 64-atom cell of diamond Si, with 0, 1 or 2 Al atoms at the T sites and all others at the Si site.

| $n$          | 1     | 2     | 4     | 8     |
|--------------|-------|-------|-------|-------|
| $n$S         | -2.89 | -2.73 | -2.57 | -2.55 |
| ($n$-1)S + 1T |       |       | -2.45 | -2.57 |
| ($n$-2)S + 2T |       |       |       | -2.56 |

Table 3 Lowest defect formation energies (in eV) versus vacuum (V) and bulk metal (B) reference states for the insertion of one atom of Li, Na and Mg at all Al concentrations studied.

|                  | Li    |       | Na    |      | Mg    |       |
|------------------|-------|-------|-------|------|-------|-------|
|                  | V     | B     | V     | B    | V     | B     |
| 0 Al             | -1.37 | 0.30  | 0.37  | 1.51 | 0.64  | 2.19  |
| 1 Al (~1.6 at. %)  | -2.12 | -0.45 | -0.42 | 0.71 | -0.46 | 1.09  |
| 2 Al (~3.1 at. %)  | -2.28 | -0.61 | -0.61 | 0.53 | -1.25 | 0.30  |
| 4 Al (~6.2 at. %)  | -2.37 | -0.70 | -0.63 | 0.51 | -1.52 | 0.03  |
| 8 Al (~12.5 at. %) | -2.43 | -0.76 | -0.75 | 0.39 | -1.78 | -0.23 |

Table 4 Energy barriers (in eV) in pure and doped Si with Al concentrations of ~1.6 and ~6.2 at. % for the diffusion of one atom of Li, Na and Mg between two T sites. $T_1$, $T_2$ and $T_3$ designate the nearest three T sites from an Al atom. $T_4$ and $T_5$ are farther T sites and are at mi-distance between two Al atoms for concentrations of ~1.6 and ~6.2 at. %, respectively.

|  | Li | | | | Na | | | | Mg | | | |
| --- | --- | --- | --- | --- | --- | --- | --- | --- | --- | --- | --- | --- |
|  | $T_1$-$T_2$ | $T_2$-$T_4$ | $T_4$-$T_5$ | $T_1$-$T_{4/5}$ | $T_1$-$T_2$ | $T_2$-$T_4$ | $T_4$-$T_5$ | $T_1$-$T_{4/5}$ | $T_1$-$T_2$ | $T_2$-$T_4$ | $T_4$-$T_5$ | $T_1$-$T_{4/5}$ |
| 0 Al | 0.56 | | | | 1.09 | | | | 0.97 | | | |
| 1 Al (~1.6 at. %) | 0.57 | 0.70 | 0.59 | 0.81 | 1.11 | 1.19 | 1.07 | 1.34 | 0.97 | 1.18 | 0.98 | 1.42 |
| 4 Al (~6.2 at. %) | 0.57 | 0.66 | | 0.72 | 1.08 | 1.11 | | 1.20 | 0.98 | 1.14 | | 1.31 |

FIGURES

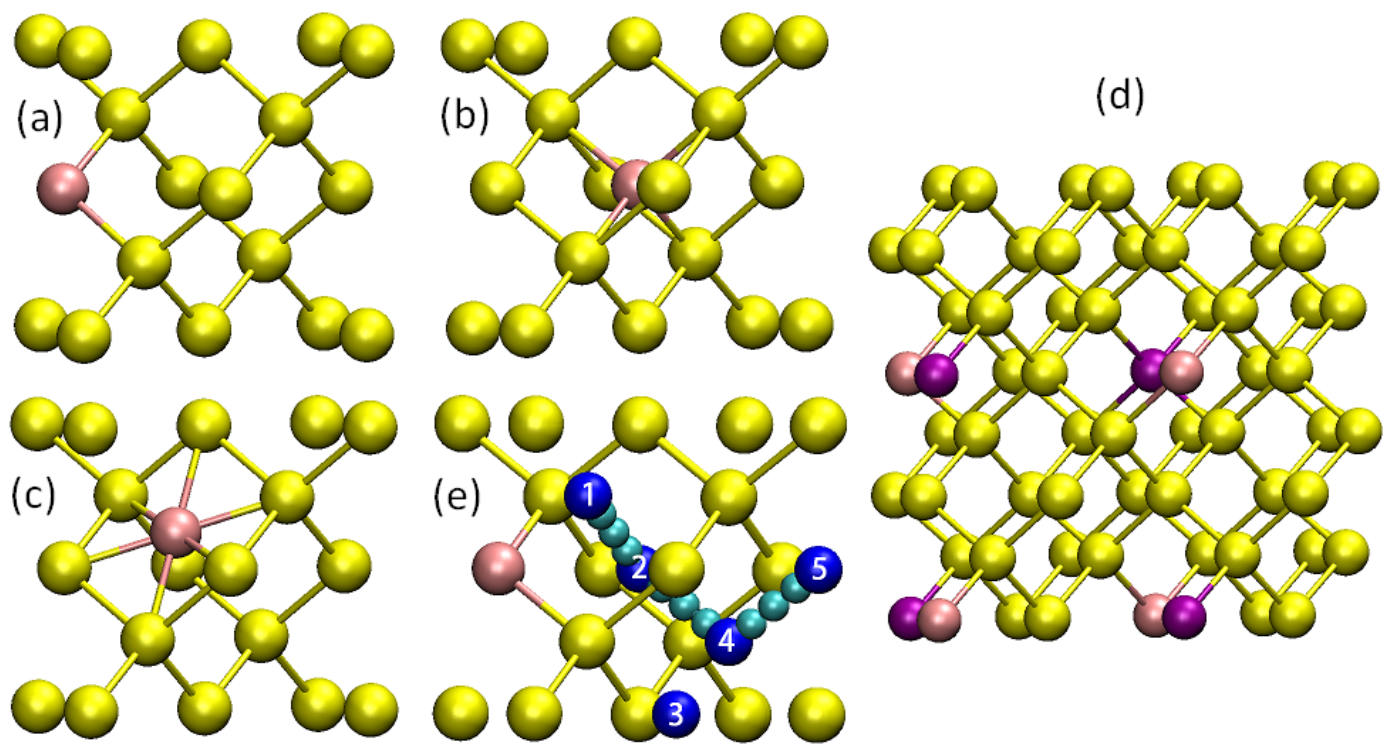

**Figure 1**

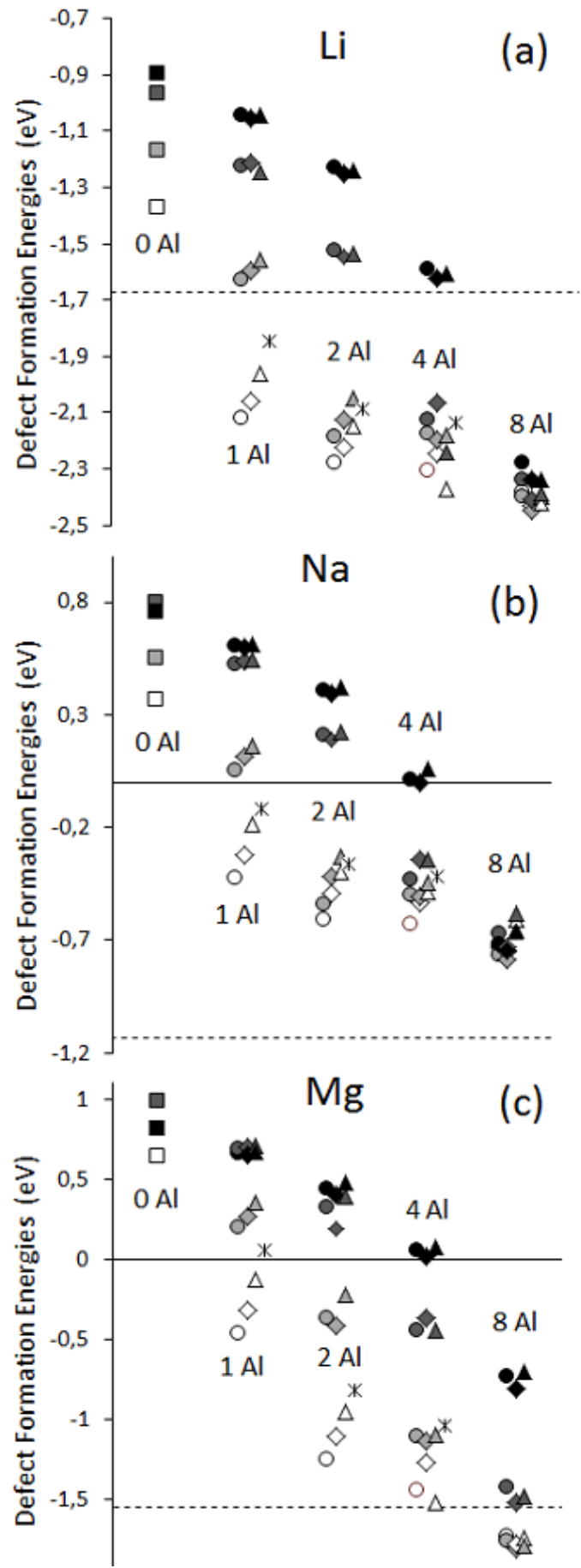

**Figure 2**

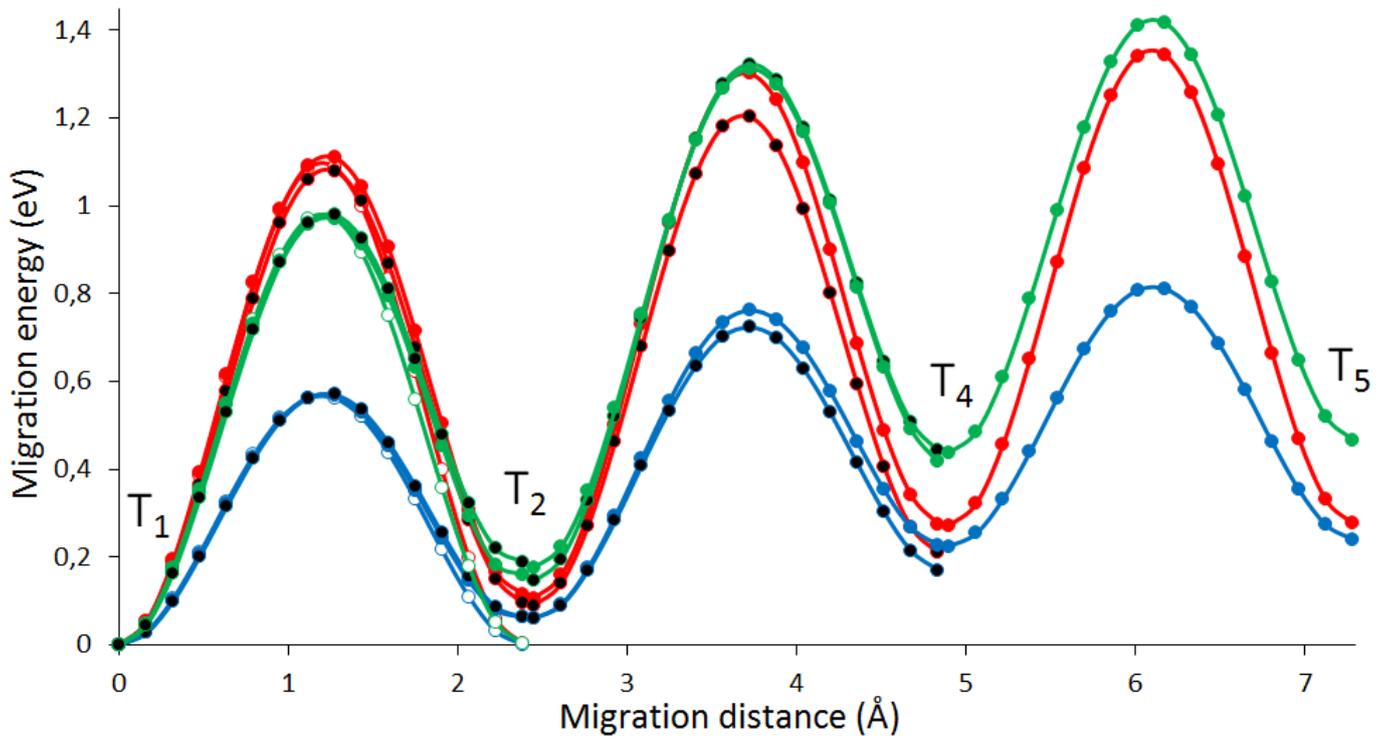

**Figure 3**

Aluminum doping improves the energetics of Li, Na, and Mg storage in silicon

**SUPPORTING INFORMATION**

*Fleur Legrain[1] and Sergei Manzhos[1]\**

[1]Department of Mechanical Engineering, National University of Singapore, Block EA #07-08, 9 Engineering Drive 1, Singapore 117576, Singapore

*\*E-mail:* mpemanzh@nus.edu.sg*; Fax: +65 6779 1459; Tel: +65 6516 4605*

### A. Permutations of Al sites

To study the insertion of alkali atoms into Al-doped Si, we considered for each Al concentration the configuration maximizing the inter-dopant distances. We investigate here how the insertion energetics changes with permutation of the Al dopant sites – by considering the S sites only - for concentrations of ~3.1 and ~6.2 at. % (i.e. for 2 and 4 Al dopants in the 64 atom cell). For all possible configurations with 2 Al dopants, 9 unique nearest inter-dopant distances exist. Their defect formation energies are plotted in Figure S1 (a). For 4 Al dopants, we characterize each configuration by its nearest inter-dopant distances among the 6 atom pairs. 313 unique sets of nearest inter-dopant distances are found. We computed one configuration for the 10 most likely cases (with the highest occurrences) as well as for the two specific configurations in which the nearest inter-dopant distances are maximized and minimized. The defect formation energies together with the occurrences of each set are plotted versus the average nearest inter-dopant distance in Figure S1 (b). Results show that for 1 (4) Al atoms, energy differences within 0.06 (0.16) eV per dopant can be induced by swapping Al sites. Figure S1 (b) also shows that Al dopants in Si tend to prefer to be clustered than well-separated.

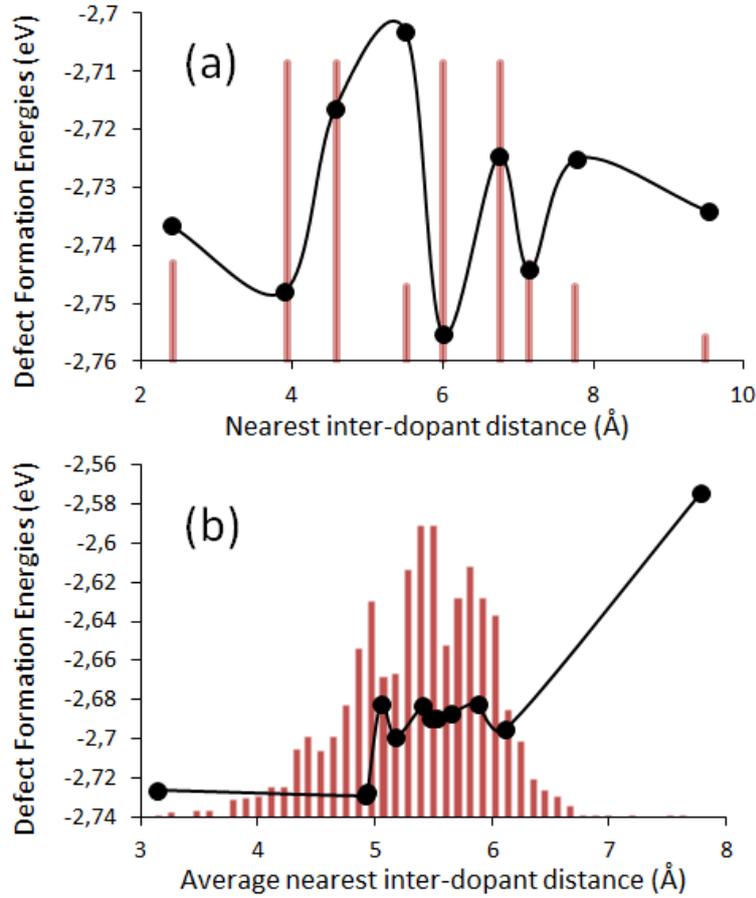

*Figure S1. (a) and (b) Defect formation energies per dopant atom versus vacuum reference state (black circles) and relative occurrence of the (average) nearest inter-dopant distance (red bars) for the insertion of 2 and 4 Al atoms in a 64 Si atom cell, respectively.*

### B. Al diffusion in Si

In the highly Al-doped Si nanowires synthesized by Moutanabbir, et al., the Al impurities were found to be homogeneously distributed in the nanowire and not form precipitates or clusters [1]. That is why we model Al-doped Si by well-separated Al dopants. However, at the Al concentrations considered, Al dopants are more stable clustered than well-dispersed. We therefore investigate the possibility for Al dopants to diffuse to form clusters. We consider three mechanisms for the diffusion of Al in Si: (i) the diffusion of Al from an (S) site to a (T) site, and then between (T) sites through (H) sites; (ii) the diffusion of Al through a vacancy; (iii) the diffusion of Al through a kick-out mechanism. These mechanisms are investigated for concentrations of ~1.6 and ~6.2 at. %.

We study the diffusion of Al from an (S) site to a (T) site by moving the Al dopant from its initial (S) site to a nearest (T) site leaving a vacancy behind. The energy needed to realize such step is found to be higher than 3.0 (2.5) eV at an Al concentration of ~1.6 (~6.2) at. %. The energy barriers corresponding to a T-H-T pathway are found to be 1.04 and 1.01 eV for ~1.6 and ~6.2 at. %, respectively. In order to study the completion of the diffusion step, in which the Al atom comes back to an S site, we also compute the Al diffusion from a T site to an S site by the kick-out mechanism. The energy barriers are found to be 1.90 (0.84) eV for concentrations of ~1.6 (~6.2) at. %.

The diffusion mechanism in which the Al dopant diffuses through a vacancy needs a vacancy to be formed near the Al atom. We model the formation of the vacancy by moving a nearest Si neighbor of Al to a T site. The energy needed to create this vacancy is found to be higher than 3 eV for the two Al concentration considered.

The kick-out mechanism is modeled by moving an Al atom towards one of its nearest Si neighbors until Al is located at the Si site and a vacancy is formed behind it. The energy barrier is found to be higher than 4 and 3 eV for Al concentrations of ~1.6 and ~6.2 at. %, respectively.

For all Al diffusion mechanisms considered, the diffusion of Al appears to be easier for a concentration of ~6.2 at. % than ~1.6 at. %, but the lowest diffusion barrier remains higher than 2.5 eV, letting the Al diffusion in Si very unlikely at these concentrations (and at room temperature). We can therefore expect the Al dopants to remain homogeneously distributed in Si, as observed by Moutanabbir et al. [1]. This justifies the configuration of well-separated Al dopants we use to model the Li, Na, and Mg storage in Al-doped Si.

### C. Strain energies

The stress induced by the insertion of Li, Na and Mg in Al-doped Si is analyzed by computing the strain energies defined as in Ref. [2]:

$$E_{strain} = E_{[64Si+M,relaxed]-M} - E_{64Si},$$

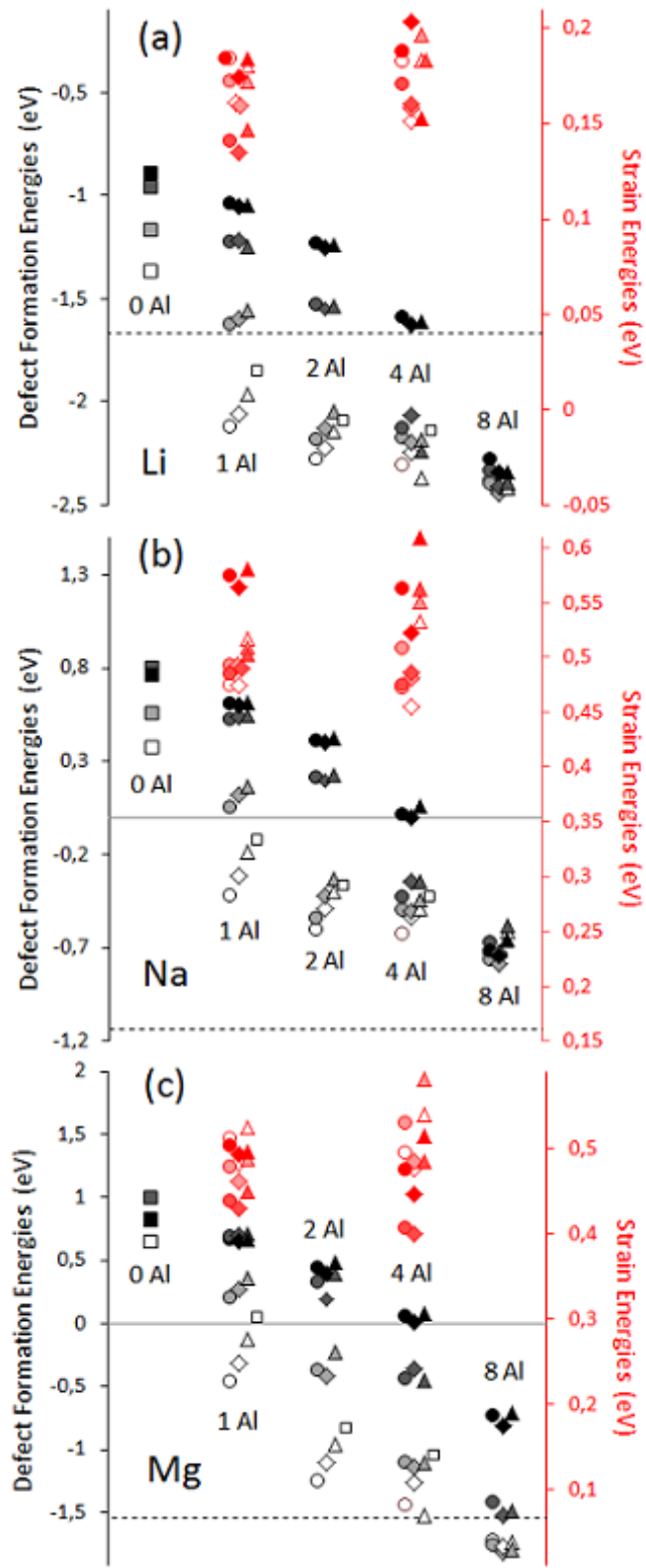

*Figure S2. (a) (b) and (c) Defect formation energies (shades of black) and strain energies (shades of red) versus vacuum reference state (in eV) for Li, Na and Mg insertion, respectively, in pure and Al-doped Si. The metal reference state is indicated with a dotted line. Symbol and color schemes: circle – $T_1$, diamond – $T_2$, triangle – $T_3$, small square – $T_f$, square – T, white – 1 M, light grey/red – 2 M, dark grey/red – 4 M, black/red – 8 M (M designating Li, Na or Mg).*

where $E_{64Si}$ and $E_{[64Si+M,\ relaxed]-M}$ represent the energies of the undoped Si supercell at the pure Si configuration and at the configuration that the Si lattice assumes in the presence of the metal atom M, respectively.

The strain energies together with the defect formation energies are plotted in Figure S2. The strain energies computed for Li, Na and Mg are in the ranges 0.13-0.20 eV, 0.46-0.61 eV and 0.41-0.58 eV, respectively. The three metals tend to generate less stress when inserted in $T_2$ rather than in $T_1$ and $T_3$ (Figure 1). Na atoms seem to generate more stress per dopant atom as the number of dopants is increased while an opposite trend appears for Mg. For 4 Al and 4 M dopants, the lowest defect formation energies are found for $T_3$ for Li and Mg, but for $T_1$ for Na. We investigate whether this difference of behavior could be explained by the stress generated by Na in $T_3$. The strain energy computed for Na in $T_3$ in such case is in fact 0.09 eV higher than in $T_1$. However, the strain energy computed for Mg in $T_3$ is also 0.08 eV higher than in $T_1$. Therefore, the strain energies are not sufficient to explain the difference in preferred sites between the different atom types.

### D. Full states of charge

We also compute the full states of charge doped with Al and computed the average voltages of the anodes [3]:

$$V = (E_{M_x(Al-doped\_Si)} - E_{(Al-doped\_Si)} - xE_M)/zx,$$

where $x$ is the number of metal M atoms per host atom, $z$ the number of electrons transferred by M atom, $E_{M_x(Al-doped\_Si)}$ is the energy of the full state of charge doped with Al, $E_{(Al-doped\_Si)}$ the energy of Al-doped Si and $E_M$ the energy of M (Li/Na/Mg) in bulk.

The voltages are given in Table S1. The results show that Al-doped and non-doped Si anodes provide very similar voltages (voltage differences within 0.03 V). Especially, the voltages given by Al-doped Si are still positive. The final states of charge will likely not change with Al doping.

*Table S1. Average anode voltages (in V) for Li, Na and Mg storage in pure and Al-doped Si (Al concentrations are given in atomic %).*

| [Al] | 0 | ~3.1 | ~6.2 | ~12.5 |
|---|---|---|---|---|
| Li | 0.20 | 0.20 | 0.20 | 0.21 |
| Na | 0.02 | 0.02 | 0.02 | 0.05 |
| Mg | 0.09 | 0.09 | 0.09 | 0.09 |

### E. Density of states

In Figure S3 are given the density of states and Fermi levels of pure and Al-doped Si. It shows that Al doping moves the Fermi level from the band gap (in pure Si) into the valence band, and the Fermi level is lowered as the Al concentration is increased.

The density of states and Fermi levels of pure and Al-doped Si with one metal atom M (Li/Na/Mg) are given in Figure S4. It shows that while the insertion of one Li/Na/Mg in pure Si moves the Fermi level into the conduction band, the presence of 1 (2) Al atoms for Li/Na (Mg) lowers the Fermi level into the band gap (and a larger Al concentration moves it into the valence band).

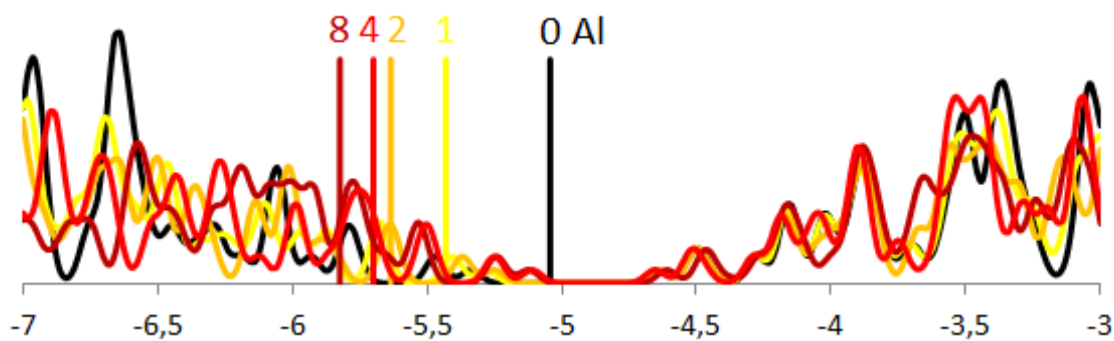

*Figure S3. Density of states and Fermi levels of pure (black) and Al-doped Si (1 Al – yellow, 2 Al – orange, 4 Al – red, 8 Al – dark red). Here and elsewhere, a broadening of 0.05 eV is used.*

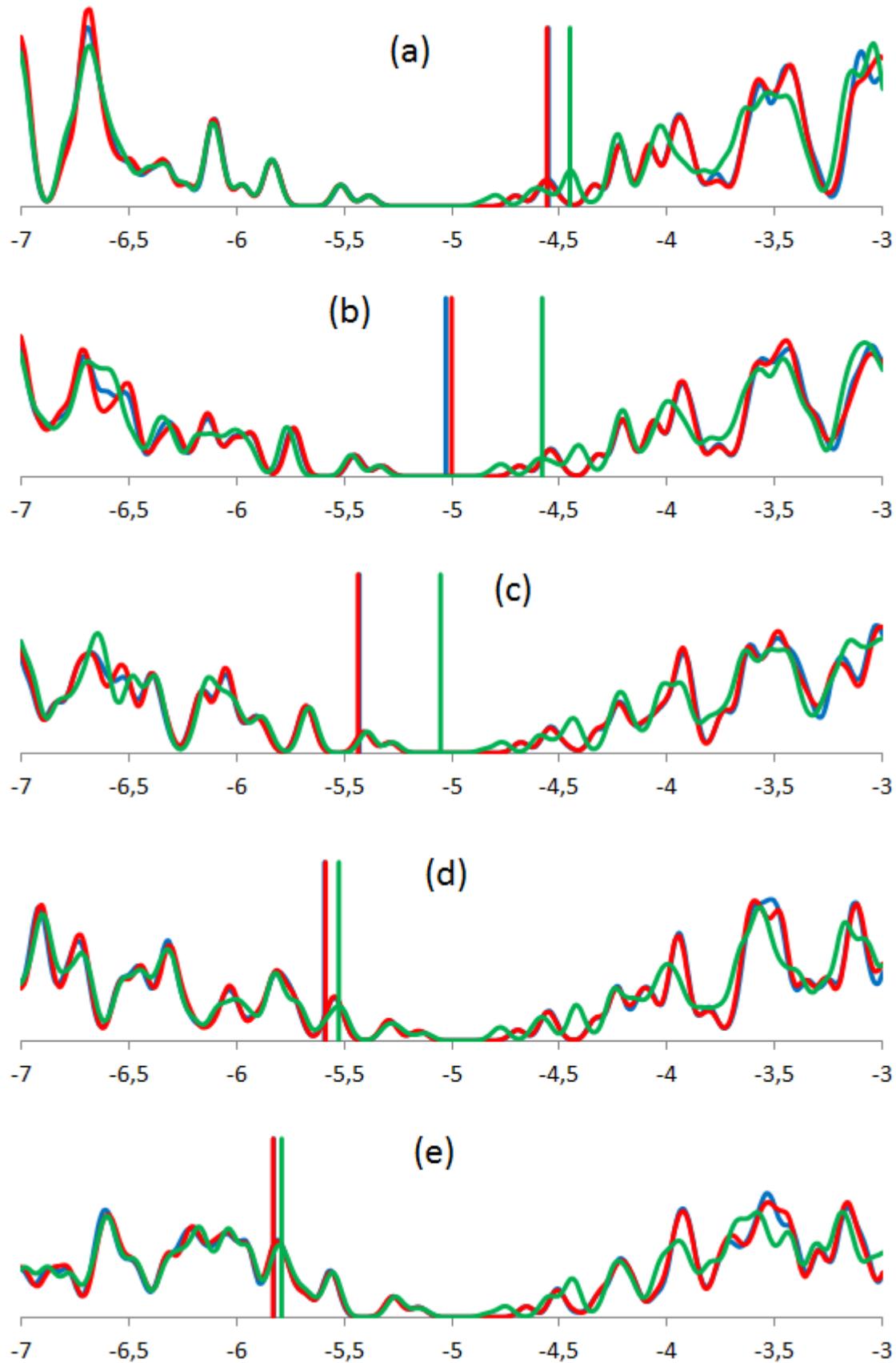

*Figure S4. (a) (b) (c) (d) and (e) show the density of states and Fermi levels (in eV) of the pure and Al-doped Si, for 0, 1, 2, 4 and 8 Al atoms, respectively, in the Si cell when one atom of Li (blue), Na (red) or Mg (green) is inserted.*

F. **Basis sets and pseudopotentials**

The basis sets and headers of the pseudopotentials used in this study are as follows:

**Si**

```
Si         2              # Species label, number of l-shells
n=3  0  2                 # n, l, Nzeta
5.965    4.531
1.000    1.000
n=3  1  2 P  1            # n, l, Nzeta, Polarization, NzetaPol
7.853    5.330
1.000    1.000
```

```
Si pb nrl pcec
ATM3     24-JUN-12 Troullier-Martins
3s 2.00  r= 1.75/3p 2.00  r= 1.94/3d 0.00  r= 2.09/4f 0.00  r= 2.09/
4 0 1074  0.177053726905E-03  0.125000000000E-01   4.00000000000
```

**Al**

```
Al         2              # Species label, number of l-shells
n=3  0  2                 # n, l, Nzeta
7.846    5.393
1.000    1.000
n=3  1  2 P  1            # n, l, Nzeta, Polarization, NzetaPol
10.859   6.753
1.000    1.000
```

```
Al pb nrl pcec
ATM3     24-JUN-12 Troullier-Martins
3s 2.00  r= 1.84/3p 1.00  r= 2.06/3d 0.00  r= 2.22/4f 0.00  r= 2.22/
4 0 1068  0.190673244359E-03  0.125000000000E-01   3.00000000000
```

**Li**

```
Li         1              # Species label, number of l-shells
n=2  0  2 P  1            # n, l, Nzeta, Polarization, NzetaPol
7.478    7.181
1.000    1.000
```

```
Li pb nrl pcec
ATM3     24-JUN-12 Troullier-Martins
2s 1.00  r= 2.26/2p 0.00  r= 2.26/3d 0.00  r= 2.59/4f 0.00  r= 2.59/
4 0  950  0.826250725555E-03  0.125000000000E-01   1.00000000000
```

**Na**

```
Na         1              # Species label, number of l-shells
n=3  0  2 P  1            # n, l, Nzeta, Polarization, NzetaPol
9.273    8.496
1.000    1.000
```

Na pb nrl pcec

ATM3     24-JUN-12 Troullier-Martins
3s 1.00  r= 2.83/3p 0.00  r= 2.83/3d 0.00  r= 3.13/4f 0.00  r= 3.13/
4  0 1054  0.225341106970E-03  0.125000000000E-01   1.00000000000

 **Mg**

Mg              1              # Species label, number of l-shells
n=3  0  2 P  1              # n, l, Nzeta, Polarization, NzetaPol
7.501    6.538
1.000    1.000

Mg pb nrl pcec
ATM3     24-JUN-12 Troullier-Martins
3s 2.00  r= 2.18/3p 0.00  r= 2.56/3d 0.00  r= 2.56/4f 0.00  r= 2.56/
4  0 1061  0.206562681389E-03  0.125000000000E-01   2.00000000000